\documentclass[a4paper]{article}
\usepackage{amsmath}
\usepackage{graphicx}
\begin{document}
\title{Where is the pseudoscalar glueball?}
\author{M. Majewski\thanks {e-mail:
m.majewski@merlin.phys.uni.lodz.pl}\\
 Department of Theoretical Physics II, University of Lodz\\
Pomorska 149/153, 90-236 Lodz, Poland\\
V. A. Meshcheryakov\\
 Bogolyubov Laboratory of Theoretical Physics, JINR,\\
Dubna Moscow Region, Russia}
\maketitle
\begin{abstract}
The pseudoscalar mesons $\pi(1300)$, $K(1460)$, $\eta(1295)$,
$\eta(1405)$ and $\eta(1475)$ are assumed to form the meson decuplet
which includes the glueball as the basis state supplementing the
standard $SU(3)_F$ nonet of light $q\bar{q}$ states $(q=u,d,s)$. The
decuplet is investigated by using the algebraic approach based on
the hypothesis of vanishing exotic commutators (VEC) of $SU(3)_F$
"charges" and their time derivatives. This leads to a system of
master equations (ME) determining: (a) octet contents of the
physical isoscalar mesons, (b) the mass formula relating all masses
of the decuplet and (c) the mass ordering rule. The states of the
physical isoscalar mesons $\eta(1295)$, $\eta(1405)$, $\eta(1475)$
are expressed as superpositions of the "ideal" $q\bar{q}$ ($N$ and
$S$) states and the glueball $G$ one. The "mixing matrix" realizing
transformation from the unphysical states to the physical ones
follows from the octet contents and is expressed totaly by the
decuplet meson masses. Among four one-parameter families of the
resulting mixing matrices (multitude of the solutions arising from
bad quality of data on the $\pi(1300)$ and $K(1460)$ meson masses)
there is a family attributing the glueball-dominated composition to
the $\eta(1405)$ meson. The pseudoscalar decuplet is similar in some
respects to the scalar one: both are composed of the excited
$q\bar{q}$ states and G; the mass ordering of their N, S, G -
dominated isoscalars is the same. Contrary to the Lattice QCD and
other predictions, the mass $m_{G^{-+}}$ of the pseudoscalar pure
glueball state is smaller than the scalar $m_{G^{++}}$ one.
\end{abstract}

\section{Introduction}
$\bullet\bullet$ The pseudoscalar glueball investigation has been
initiated soon after it was realized that bound states of the gluons
may play an important role in the strong interactions \cite{Fri-Gel,
Fri-Min}. From the very beginning the glueball state was traced
within structure of the $\eta$ and $\eta'$ mesons \cite{Ros, Gov}.
At present, this is not the main purpose of the investigation, but
is still continued, and not only within the meson structures
\cite{Kou, KLOE} but also within the baryon ones \cite{Maj-Mes}.

The discovery of the $\iota$ meson \cite{Sch, Edw} rous hopes for
the existence of the glueball. The $\iota$ meson has been detected
in the gluon rich process of the $J/\psi$ radiative decay and was
immediately claimed to be a glueball. However, the glueball may
exist as separate particle only if it has exotic quantum numbers;
otherwise it should be mixed with the isoscalar $q\bar{q}$ states
having the same signatures $J^{PC}$. The mass of the $\iota$ meson
belongs to the region of a higher-lying $0^{-+}$ multiplet and the
states of this multiplet were (and still are) poorly known. That
posed the question of how to certify such assignment. To this end
several criteria have been invented which could be used in the cases
of deficient multiplets. Some of them, concerning production, are
pure qualitative like "creation in the gluon-rich environment",
other ones, more regarding decay products, are semi-quantitative
(big value of "stickness" \cite {Cha} and "gluiness"
\cite{Clo-Far-Zhe}). At the same time, the question has been risen
whether the glueball is necessary for understanding data concerning
the pseudoscalar mesons known at that time \cite{GG1}. This question
is still alive \cite{Kle}.

The trend of discussion has changed since the results of the Lattice
QCD (LQCD) calculations became available \cite{Bali, Mor, Che}. They
supported the very existence of the pseudoscalar glueball, but the
mass attributed was about $2.3 GeV$ - much above the $\iota$. An
attempt to lower the lattice prediction by including quark loops was
not very successful \cite{Gab}. Although the doubts were not
dispelled (see e. g. \cite{MO}), this became a serious obstacle for
$\iota$ to be recognized as the  glueball candidate, because the
results of the lattice calculations are generally accepted. On the
other hand, there is no candidate having the mass predicted by
lattice. Perhaps that induced several attempts to interpret the
meson $X(1835)$ as the pseudoscalar glueball \cite{Koc-Min, Bin,
Tao}, although its mass  might also be regarded as too low.

At the same time, starting from late 1980s there was growing
conviction that the $\iota(1440)$ signal should be attributed to two
different isoscalar mesons \cite{Lip}. Much experimental effort was
devoted to understanding the structure of the signal \cite{Bai, Acc,
Aug, Bol, Ams}. As a result, it has been split into $\eta(1405)$ and
$\eta(1475)$. Hence, since 2004 three isoscalar pseudoscalar mesons
have been listed in RPP within the narrow interval of mass
\cite{PDG}:
\begin{equation}\label{1.1}
 \eta_1=\eta(1295),\quad  \eta_2=\eta(1405),\quad  \eta_3=\eta(1475).
\end{equation}
Such three isoscalar mesons with similar masses in the vicinity of
the isotriplet and isodublet suggest overpopulation of a nonet and
possible existence of a glueball which is hidden within the
structures of three isoscalar states. The decuplet findings are very
important because investigation of its properties is the most
promising way of glueball search unless the glueball with exotic
quantum numbers will be detected.

 Information about the structures of the isoscalar
mesons $\eta_1$, $\eta_2$, $\eta_3$ has been extracted from data on
the reactions of their production and from branching ratios of their
decays. Data suggest that the meson $\eta_2$ is a particle dominated
by the glueball state \cite{Clo-Far-Zhe, PDG, Mas-Cic-Usa, Li}. \\

 $\bullet\bullet$ An unexpected objection has been risen against such a picture: the
$\eta_1$ has been claimed not to be the $q\bar{q}$ state \cite{Kle}.
Even its very existence was considered uncertain. This implies the
nonexistence of the decuplet and requires much more complicated
spectroscopy of the pseudoscalar mesons. Therefore, we discuss this
question in more detail.

The $q\bar{q}$ structure is put into doubt due to not occurrence of
the $\eta_1$ in the reactions
\begin{equation}\label{1.2}
    p\bar{p}, \quad J/\psi, \quad \gamma\gamma \rightarrow\quad,
\end{equation}
"at least not with the expected yields". The base of such
expectation is not indicated.

However, this is not the only point of view concerning  these
reactions. The authors of recently published, very careful analysis
of the experimental data on $\eta_1$, came to the following
conclusions \cite{Mas-Cic-Usa}:\\
(i) the charge exchange experiments $\pi^--p \rightarrow
~n\eta\pi\pi,\quad nK\bar{K}\pi^0$ definitively establish evidence
of the $\eta_1$;\\
(ii) a clear signal of $\eta_1$  is seen in the $J/\psi$ radiative
decay;\\
(iii) there is an indication for the existence  $\eta_1$ in $p\bar p$ annihilation;\\
(iv) the LEP data on $\gamma\gamma$ reaction are compatible with the
existence of the $\eta_1$ signal.

So the $\eta_1$ may or may not be seen in the reactions (\ref{1.2}).
The three-body decays $\eta_1 \rightarrow\eta\pi\pi\ $,
$K\bar{K}\pi$ are strongly suppressed by small phase space (cf
$\omega\rightarrow\pi\pi\pi)$ and that may be the reason why it is
difficult to observe the $\eta_1$. It is explicitly seen in the
reaction $\pi^--p \rightarrow ~\eta_1 n$, for which high statistics
is available, but the number of events observed in reactions
(\ref{2}) is many times smaller \cite{PDG}. Obviously, more
measurements are needed to elucidate the situation. But this
question has no relevance to the problem of the $\eta_1$ internal
structure. As in the reaction (\ref{2}) the $\eta_1$ is observed
throughout the products of decay, the frequency of its registration
depends on width; the subsequent measurements would verify the
magnitude of the width. However, the definition of the multiplet
does not depend on the widths. Therefore, the widths of the
particles cannot be the basis for any conclusion about the structure
of the multiplet. Also the width of the $\eta_1$ cannot be the base
for conclusion about its $q\bar{q}$ structure. An attempt to call in
question this structure resembles confusion which arose after
denying the $q\bar{q}$ structure of the $f_0(980)$ meson motivated
by its small width \cite{enc}.

 In the present paper we admit the $\eta_1$ meson to be a
$q\bar{q}$ state and assume that the examined pseudoscalar mesons
form a decuplet. We thus focus the glueball search again in the
region of $\iota$ meson - this time being fully aware of the
conflict with the lattice prediction.

The glueball assignment of the $\eta_2$ meson is also motivated on
theoretical ground \cite {Fad}. It is argued that $\eta_2$ meson is
a natural pseudoscalar glueball candidate if the $f_0(1500)$ is the
scalar glueball and the glueballs are described as the closed
gluonic flux-tubes. Then the $f_0(1500)$ and the $\eta(1405)$ would
be two parity related glueballs with equal masses.  This description
deserves attention in view of failure of the Lattice prediction,
particularly if it can be treated not too literally.

It is thus interesting to make sure that this assignment can be
confirmed by an argument based on the properties of the flavor
multiplet as a whole.\\

$\bullet\bullet$ We conclude the introduction with few comments
concerning credibility of the approach we use in this paper. The
credibility is especially important in evaluating the glueball
contents of the decuplet isoscalar states.

It is currently known that broken $SU(3)_F$ symmetry predicts the
existence of octets and nonets of light mesons. The multiplets are
usually testified by the mass formula relating their masses. The
Gell-Mann--Okubo (GMO) and Schwinger (S) mass formulae have been
obtained by inclusion into the lagrangean the non-invariant mass
term with regard to mixing of the octet isoscalar with the unitary
singlet.

Our model unifies and generalizes these mass relations. The model
has been introduced at the University of Lodz in the middle of '80s
\cite{MT1, MT} and is based on requirement of vanishing the exotic
commutators (VEC) of the "charges" and their time derivatives. Apart
from the GMO and S mass formulae it gives additional insight into
the properties of the multiplet. For the S nonet the model VEC
determines the mixing angle and establishes the mass ordering rule
which ensures the mixing angle to be a real number. There are two
possible orderings. For one of them the mixing angle $\vartheta$ is
smaller than ideal $\vartheta<\vartheta^{id}$, while for the other
one it is bigger $\vartheta>\vartheta^{id}$
($\vartheta^{id}\approx35^o$).

The model also predicts the ideally mixed (ideal) nonet (I). This
nonet has not been derived, as yet, from any other mixing
description. In the quark model, where it is the basic object, it is
postulated.

The S and I mass formulae are well obeyed by many nonets with
various signatures $J^{PC}$ comprising low mass mesons \cite{Sum}.
In general, the S nonets better describe data, although differences
between I and S descriptions are small.

For the glueball quest the most important is the prediction of a
decuplet \cite{MT} - a multiplet comprising three isoscalar mesons.
The mass formula, mass ordering rule and the octet contents of the
physical isoscalar states follow from the same constraints.  The
octet contents $l_i^2$ play a key role in determining mixing matrix
of the isoscalar states, i.e. the contributions of glueball state to
their structures. The orthogonal 3 x 3 mixing matrix can be
parametrized by Euler angles. Absolute values of the trigonometric
functions of these angles are expressed by the particle masses.

The VEC model does not use additional assumptions nor introduces
free parameters to describe multiplets. Its predictions are definite
and applicable for decuplets of any signature $J^{PC}$. If fitted
with required experimental input, it offers complete description of
the decuplet states. Thus, it bestows quantitative meaning to the
most obvious qualitative signature of the glueball presence --
overpopulation of a nonet.

The model has appeared very effective in describing the $0^{++}$
mesons. It makes possible to sort out 20 scalar mesons among
multiplets and attribute the glueball dominating structure to the
$f_0(1500)$ meson \cite{enc}. To analyze the $0^{-+}$ decuplet we
use essentially the same model. Our present analysis is proceeded in
a different way because sample of the input data is different.

\section{The decuplet of pseudoscalar mesons}
\subsection{The model of vanishing exotic commutators (VEC)\protect\footnotemark}

The following sequence of exotic commutators is assumed to vanish
\footnotetext{formerly the model was called exotic commutator model
(ECM)}
\begin{equation}
    \left[T_a,\frac{d^{j}T_b}{dt^j}\right]=0,\quad \left(j=1,2,3,...\right)
\label{2.01}
\end{equation}
where $T$ is $SU(3)_F$ generator, $t$ is the time and $(a,b)$ is an
exotic combination of indices, i.e. such that the operator
$[T_a,T_b]$ does not belong to the octet representation.
Substituting $\frac{dT}{dt}=i[H,T]$, and using the infinite momentum
approximation for one-particle hamiltonian $H = \sqrt{m^2+p^2}$
\cite{Tyb}, we transform eqs.~(\ref{2.01}) into the system:

\begin{displaymath}
    [T_a,[\hat{m^2},T_b]]=0,
\end{displaymath}
\begin{displaymath}
    [T_a,[\hat{m^2},[\hat{m^2},T_b]]]=0,
\end{displaymath}
\begin{equation}
    [T_a,[\hat{m^2},[\hat{m^2},[\hat{m^2},T_b]]]]=0, \label{2.02}
\end{equation}
\begin{displaymath}
........................................
\end{displaymath}
where $\hat{m^2}$ is the squared-mass operator.

For the matrix elements of the commutators (\ref{2.02}) between
one-particle states (we assume one-particle initial, final and
intermediate states) we obtain the sequence of equations involving
expressions $\langle x_8|(m^2)^j|x_8\rangle$ with different powers
$j=1,2,3,..$, where $x_8$ is the isoscalar state belonging to the
octet. Solving these equations, we obtain the sequence of formulae
for a multiplet of the light mesons. We find
\begin{equation}
    \langle{x_8}\mid{\hat{(m^2)}^j}\mid{x_8}\rangle = \frac{1}{3} a^j +
    \frac{2}{3}b^j \quad (j=1,2,3,...). \label{2.03}
\end{equation}
where $a$ is the mass squared of the isovector meson $\pi$; $b$ is
the mass squared of the subsidiary $s\bar{s}$ state,
\begin{equation}
    b=2K-a,    \label{2.04}
\end{equation}
and $K$, in turn, is the mass squared of the isospinor $K$ meson.
The isoscalar octet state $\mid{x_8}\rangle$ can be represented as
the linear combination of the physical isoscalar states
\begin{equation}
 \mid{x_8\rangle}=\sum{{l_i\mid{x_i}\rangle}}.
 \label{2.05}
\end{equation}
The coefficients $l_1$, $l_2$, $l_3$,.. determine octet contents of
the physical isoscalar states $|x_1\rangle$ $|x_2\rangle$,
$|x_3\rangle$,...  Substituting (\ref{2.05}) into (\ref{2.03}) we
obtain \textit{master equations} (ME) of the multiplet.
\begin{equation}
    \sum{l_i^2x_i^j}=\frac{1}{3}a^j+\frac{2}{3}b^j, \quad (j=0,1,2,3,...)
    \label{2.06}
\end{equation}
where the $x_1$, $x_2$, $x_3$,... are isoscalar meson masses
squared. Normalization condition of the $l_i$ coefficients is
included into (\ref{2.06}) as equation for $j=0$.

\subsection{ME for the decuplet}
The states of the decuplet belong to a reducible representation of
the $SU(3)_F$
\begin{equation*}
    \bf{8\oplus1\oplus1},\rm
\end{equation*}
where the octet and one of the singlets are considered as $q\bar{q}$
states while the second singlet is supposed to be a glueball $G$.

For the decuplet we have following system of the ME \cite{MT, MT1}
which determines masses and mixings of the decuplet states
\cite{enc}:
\begin{equation}
    l_1^2x_1^j+ l_2^2x_2^j+l_3^2x_3^j=\frac{1}{3}a^j+\frac{2}{3}b^j, \quad (j=0,1,2,3)
    \label{2.1}
\end{equation}
where $x_1, x_2, x_3$ are the masses squared of the isoscalar mesons
$\eta_1$, $\eta_2$, $\eta_3$.\\
The coefficients $l_1$, $l_2$, $l_3$, are real, as all isoscalar
mesons are neutral particles. The ME (\ref{2.1}) are considered as a
system of linear equations with respect to unknown coefficients
$l_i^2$.

The solution is given by three kinds of relations \cite{MT, enc}:\\
a) the octet contents (OC) of the isoscalar states
\begin{subequations}\label{2.5}
\begin{align}
\label{2.5a}
    l_1^2=\frac{1}{3}\frac{(x_2-a)(x_3-a)+2(x_2-b)(x_3-b)}
    {(x_1-x_2)(x_1-x_3)}, \\
\label{2.5b}
    l_2^2=\frac{1}{3}\frac{(x_1-a)(x_3-a)+2(x_1-b)(x_3-b)}
    {(x_2-x_1)(x_2-x_3)}, \\
\label{2.5c}
    l_3^2=\frac{1}{3}\frac{(x_1-a)(x_2-a)+2(x_1-b)(x_2-b)}
    {(x_3-x_1)(x_3-x_2)};
\end{align}
\end{subequations}
 b) the mass formula (MF)
\begin{equation}
    f(a)+2f(b)=0, \label{2.6}
\end{equation}
where
\begin{equation}\label{2.7}
 f(x)=(x_1-x)(x_2-x)(x_3-x)
    \end{equation}
is the characteristic polynomial of the $m^2$ operator; the
numbering of its eigenvalues is chosen such as to satisfy the
inequality
\begin{equation}\label{2.8}
   x_1<x_2<x_3;
\end{equation}
c) the mass ordering rule (MOR)
\begin{equation}\label{2.9}
    x_1<a<x_2<b<x_3.
\end{equation}
The MF (\ref{2.6}) is a linear equation with respect to each of the
$x_i$, but it is a cubic one with respect to $a$ and $b$.

The masses and experimental status of the mesons assigned to the
decuplet are quoted in tab. 1.  The table shows that masses of the
isoscalar mesons $\eta_1$, $\eta_2$, $\eta_3$ are determined with
good accuracy; the $\pi(1300)$ meson mass has large error; the
$K$-meson is not yet established -- its mass is unknown. Therefore,
these two masses should be considered unknown. It is natural in this
model to choose $a$ and $b$ (\ref{2.04}) as unknown variables of the
ME.
\begin{table}
 \centering
 \caption{Pseudoscalar mesons merged into decuplet.
  Status of the particles and their masses (in MeV) are quoted after RPP
  \cite{PDG}}\label{1}
  \smallskip
\begin{tabular}{|c|c|c|c|c|}
  \hline
  $\bullet\pi(1300)$ &$K(1460)$ &$\bullet\eta(1295)$ &$ \bullet\eta(1405)$ &$ \bullet\eta(1475)$ \\
  \hline
  $1300\pm100$ &  & $1294\pm4$ & $1410.3\pm2.6 $& $1476\pm4$ \\
  \hline
\end{tabular}
\end{table}

For solving the ME and constructing the mixing matrix of the
decuplet the solution of the MF is needed. However, MF is a single
equation and its solution cannot be unique. Yet, high precision of
the data on isoscalar meson masses provides correct form of the
characteristic polynomial f(x) of the $m^2$ operator as well as
precise values of the $a$ and $b$ bands which are required by MOR
(\ref{2.9}). We hope that restrictions of the model on decuplet
states will reduce ambiguity of the solution.

The restrictions imposed by MOR (\ref{2.9}) are obvious. The unknown
variables $a$ and $b$ have to satisfy MOR  which requires $a\in(x_1,
x_2)$ and $b\in(x_2, x_3)$. That restricts also the K-meson mass. It
follows from (\ref{2.04}) and (\ref{2.9}) that
\begin{equation}\label{2.10}
    x_1+x_2<2K<x_2+x_3,
\end{equation}
or
\begin{equation}\label{2.11}
    1353 MeV<m_K<1443 MeV.
\end{equation}
From $a\in(x_1,x_2)$ we have
\begin{equation}\label{2.12}
    1294 MeV<m_\pi<1410 MeV.
\end{equation}

Comparing the bands (\ref{2.12}) with the range of error of the
$\pi(1300)$ meson mass we find that the MOR cuts off lower part of
the error range and that the MOR-allowed region covers the upper
part of it. This is consistent with treating $a$ as an unknown
quantity of the ME. \\

\subsection{Families of solutions of the ME}
Combining the MOR with MF we restrict the unknown masses much
stronger. Moreover, as will be seen below, the allowed masses can be
attributed to the solutions of ME with explicit flavor properties.

Figure 1 displays $f(x)$ and $-2f(x)$ (\ref{2.6}) (c.f. \cite{KM}).
The pair of the unknown variables ($a,b$) provides a solution of the
MF if they are such that $f(a)=-2f(b)$.
\begin{figure}
\centering
\includegraphics[angle=-90,width=\textwidth]{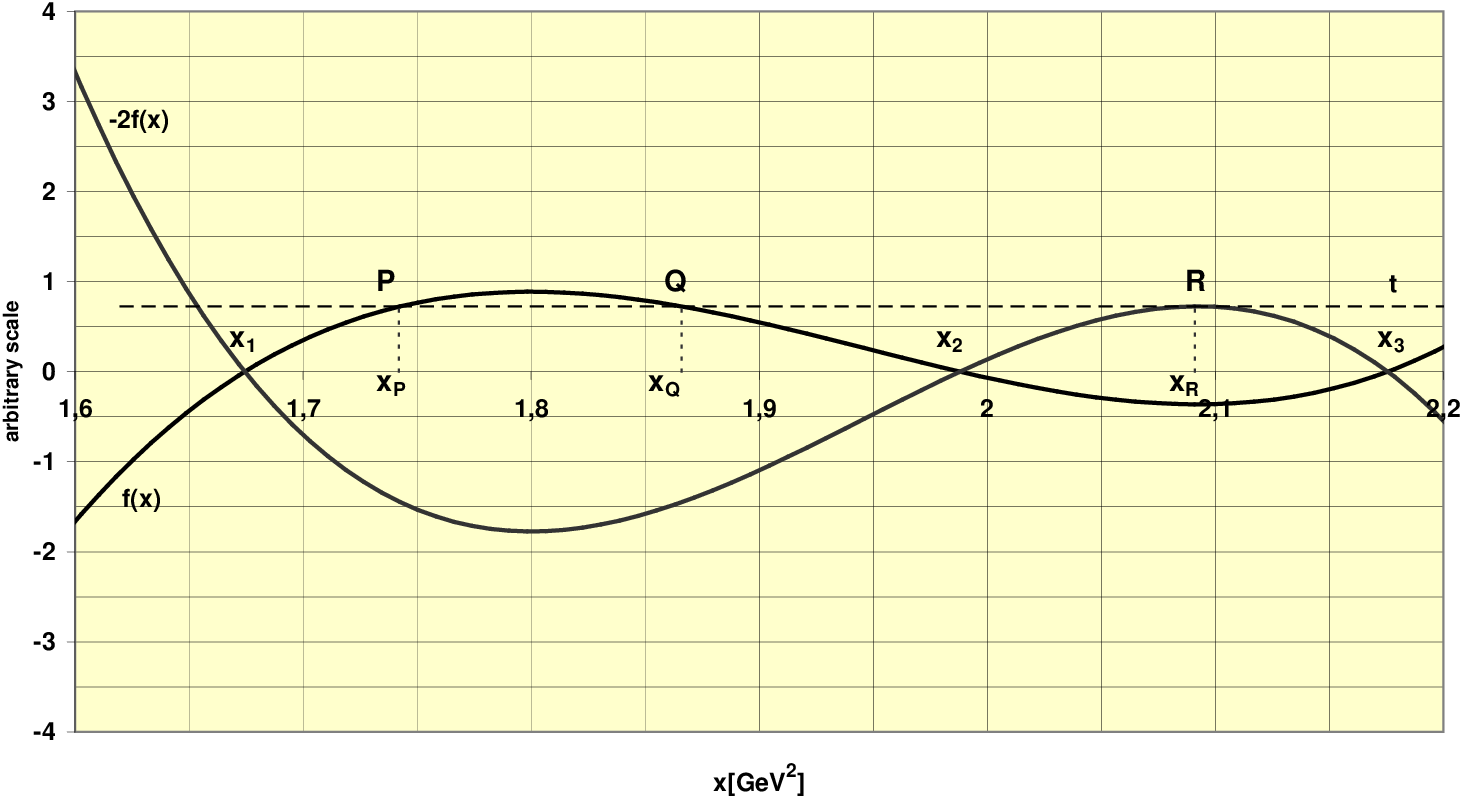}
\caption{The allowed values of the unknown quantities $a$ and $b$.
The function $f(x)$ (\ref{2.7}) is the characteristic polynomial of
the $m^2$ operator. The eigenvalues $x_1, x_2, x_3$ are squared
masses of the physical isoscalar mesons $\eta(1295)$, $\eta(1405)$,
$\eta(1475)$. The function $-2f(x)$ is also shown. The $a$ and $b$
are restricted by ordering rule ($\ref{2.9}$) and related by the
mass formula (\ref{2.6}): $f(a)=-2f(b)$. The horizontal line $t$
which is tangent to the curve $-2f(x)$ at the  point $R$ of the
local maximum ($x_R\in(x_2,x_3)$) crosses the curve $f(x)$ at the
points P and Q. The figure indicates that the MF cannot be satisfied
for $a\in(x_P,x_Q)$.} \label{Figure1}
\end{figure}
It can be seen from the figure that beside the MOR restrictions
$a\in(x_1, x_2)$, $b\in(x_2, x_3)$ there also appears the MF
restriction forbidding $a\in(x_P, x_Q)$. Hence, the allowed values
of $a$ belong to two narrow intervals: $a\in(x_1, x_P)$ and
$a\in(x_Q, x_2)$. To each allowed value of $a$ there correspond two
values of $b$ (obeying the mass formula) placed on the opposite
sides of the point $x_R$. If we wish to have unique solution, we
should divide the interval $(x_2, x_3)$ into two parts: $(x_2, x_R)$
and $(x_R, x_3)$. Then, we get four domains including unique pairs
of values $(a,~b)$ making solutions of the MF:
\begin{subequations}\label{2.13}
\begin{align}
    \label{2.13a}
    A:&&   a&\in(x_1, x_P),&    b&\in(x_R, x_3),\\
    \label{2.13b}
    B:&&   a&\in(x_1, x_P),&    b&\in(x_2, x_R),\\
    \label{2.13c}
    C:&&   a&\in(x_Q, x_2),&    b&\in(x_R, x_3),\\
    \label{2.13d}
    D:&&   a&\in(x_Q, x_2),&    b&\in(x_2, x_R),
\end{align}
\end{subequations}
where
\begin{equation}\label{2.14}
    x_P\simeq(1.320GeV)^2,\quad x_Q\simeq(1.365GeV)^2,\quad
    x_R\simeq(1.447GeV)^2.
\end{equation}
These values correspond to
\begin{equation}\label{2.14a}
    x_1=(1.294GeV)^2,\quad x_2=(1.410GeV)^2,\quad x_3=(1.475GeV)^2.
\end{equation}

The domains A, B, C, D are shown in figure 2. We solve the ME
(\ref{2.1}) in each of them separately and express $b$  as functions
of $a$. The details of solving the MF as well as properties of the
solutions are described in the appendix.

Next we calculate the octet contents of the physical isoscalar
decuplet states $l_1^2$, $l_2^2$, $l_3^2$ (\ref{2.5}) and construct
the mixing matrix of the states $\eta_1$, $\eta_2$, $\eta_3$
\cite{enc}.
The mixing matrix V is chosen such that\\
\begin{equation}\label{2.20}
           \begin{bmatrix}
         \eta_1 \\
         \eta_2 \\
         \eta_3 \\
       \end{bmatrix}
     =V\begin{bmatrix}
          N \\
          S \\
          G \\
        \end{bmatrix},
\end{equation}
where
\begin{equation}\label{2.21}
    N=\frac{1}{\sqrt{2}}(u\bar{u}+d\bar{d}), \qquad S=s\bar{s},\qquad G
- glueball.
\end{equation}\\
So V expresses the states of the physical isoscalar mesons $\eta_1$,
$\eta_2$, $\eta_3$ in terms of the decuplet ideal states $N$, $S$
and $G$. The V is an orthogonal matrix. Its elements are defined by
the masses.

In each of the domains A, B, C, D there is one point where the
solution of the ME (\ref{2.1}) is degenerate. The points are placed
at the corners of domains A, B, C, D as is shown in Fig. 2. In the
first three domains we find three different ideally mixed $q\bar{q}$
nonets and a detached glueball; in the domain D we obtain degenerate
decuplet composed of the octet states and two singlet states
detached from the octet:
\begin{subequations}\label{2.18}
\begin{align}
    \label{2.18a}
    A:&&   x_1&=a, \quad x_3=b,\quad l_1^2=\frac{1}{3},\quad l_2^2=0,\quad l_3^2=\frac{2}{3},\\
    \label{2.18b}
    B:&&   x_1&=a, \quad x_2=b,\quad l_1^2=\frac{1}{3},\quad l_2^2=\frac{2}{3},\quad  l_3^2=0,\\
    \label{2.18c}
    C:&&   x_2&=a, \quad x_3=b,\quad l_1^2=0,\quad l_2^2=\frac{1}{3},\quad l_3^2=\frac{2}{3},\\
    \label{2.18d}
    D:&&   x_2&=b=a=x_8, \quad l_1^2=0,\quad l_2^2=1,\quad l_3^2=0.
\end{align}
\end{subequations}
For the wave functions one obtains:
\begin{subequations}\label{2.19}
\begin{align}
    \label{2.19a}
    A:&&   \eta_1&=\pm N, \qquad \eta_2=\pm G, \quad \eta_3=\pm S,\\
    \label{2.19b}
    B:&&   \eta_1&=\pm N, \qquad \eta_2=\pm S, \quad \eta_3=\pm G,\\
    \label{2.19c}
    C:&&   \eta_1&=\pm G, \qquad \eta_2=\pm N, \quad \eta_3=\pm S,\\
    \label{2.19d}
    D:&&   \eta_1&=~~\gamma_1,\qquad \eta_2=\pm x_8,\quad
\eta_3=\gamma_2.
\end{align}
\end{subequations}
Each of the degenerate solutions A, B, C points out its own
candidate from among $\eta_1$, $\eta_2$, $\eta_3$ as a pure
glueball. The solution D describes degenerate decuplet where
$\eta_2$ is the octet isoscalar $\eta_8$ state and $\eta_1$,
$\eta_3$ are scalar states built as superpositions of the
$(q\bar{q})_{singlet}$ and G.

The intervals $(x_1, x_P)$ and $(x_Q, x_2)$ of the variable $a$
allowed by MF and MOR are small. Also the intervals $(x_2,x_R)$ and
$(x_R,x_3)$ of the variable $b$ are small. Therefore, the domains A,
B, C, D are also small and across any domain the solutions are not
much different from the degenerate ones. The solutions of ME in any
given domain constitute one-parameter family. To each of the domains
there corresponds such a family. The solutions belonging to the same
family are dominated by the same structure (N, S, G, $\eta_8$) which
is pure in the degenerate solutions. That can be seen from the Tab.
2. Hence, the dominant structures of the $\eta_1$, $\eta_2$,
$\eta_3$ in the domains A, B, C, D preserve the patterns of
degenerate decuplet (\ref{2.19}):
\begin{subequations}\label{2.22}
\begin{align}
    \label{2.22a}
    A:&&   \eta_1&\sim N, \qquad \eta_2\sim G, \qquad \eta_3\sim S,\\
    \label{2.22b}
    B:&&   \eta_1&\sim N, \qquad \eta_2\sim S, \qquad \eta_3\sim G,\\
    \label{2.22c}
    C:&&   \eta_1&\sim G, \qquad \eta_2\sim N, \qquad \eta_3\sim S,\\
    \label{2.22d}
    D:&&  \eta_1&\sim\gamma_1,\qquad \eta_2\sim \eta_8,\qquad
\eta_3\sim\gamma_2,
\end{align}
\end{subequations}
where $\eta_8$ is the octet isoscalar state and $\gamma_1$,
$\gamma_2$ are superpositions of the $q\bar{q}$ singlet and G. Their
contribution to the $\gamma_1$ and $\gamma_2$ states can be
expressed by masses of the physical isoscalar mesons and are slowly
varying functions inside the domain D.

Let us give the examples of the mixing matrix of the A, B, C, D
solutions near degeneracy.

In each example the value of parameter $\Delta{a}$ is chosen such
that the deviation of the $\pi(1300)$ meson mass $m_a$ from its
ideal value (i. e. from the $\eta_1$ or from the $\eta_2$ meson
mass) is equal to $6 MeV$. The choice of this number is to some
extend arbitrary. We want to have a decuplet which is deviated both
not too little and not too much from the degenerate one; 6 MeV is
the difference between the mean RPP values of $\pi(1300)$ and
$\eta_1$
masses. \\

A. $a=(1.300GeV)^2$,  $b=(1.471652GeV)^2$,  $K=(1.388GeV)^2$,\\
  $l_1^2=0.297415$, \quad $l_2^2=0.112682$, \quad $l_3^2=0.589903$,
\begin{equation}\label{2.23}
    V_A=\left(
          \begin{array}{ccc}
            0.975488 & 0.021850 & 0.218966 \\
            0.217116 & -0.257600 & -0.941543 \\
            0.035833 & 0.966004 & -0.256030 \\
          \end{array}
        \right);
 \end{equation}

B. $a=(1.300GeV)^2$,  $b=(1.414537GeV)^2$,  $K=(1.358GeV)^2$,\\
 $l_1^2=0.297415$,  \quad  $l_2^2=0.689645$, \quad
$l_3^2=0.012941$,
\begin{equation}\label{2.24}
    V_B=\left(
          \begin{array}{ccc}
            0.991872 & 0.033435 & 0.122771 \\
            0.059285 & -0.975166 & -0.213391 \\
            0.112588 & 0.218935 & -0.969222 \\
          \end{array}
        \right);
\end{equation}

C. $a=(1.404GeV)^2$,  $b=(474088GeV)^2$,  $K=(1.439GeV)^2$,\\
$l_1^2=0.002985$,  \quad $l_2^2=0.374721$,  \quad $l_3^2=0.622294$,
\begin{equation}\label{2.25}
    V_C=\left(
          \begin{array}{ccc}
            0.233852 & 0.098445 & 0.967276 \\
            0.971521 & -0.062752 & -0.228491 \\
            0.038204 & 0.993162 & -0.110316 \\
          \end{array}
        \right);
\end{equation}

D. $a=(1.404GeV)^2$,  $b=(1.413137GeV)^2$,  $K=(1.409GeV^2$,\\
$l_1^2=0.000598$,  \quad $l_2^2=0.996963$, \quad $l_3^2=0.002439$,
\begin{equation}\label{2.26}
    V_D=\left(
          \begin{array}{ccc}
            0.530736 & 0.345334 & 0.773992 \\
            0.595077 & -0.802101 & -0.050176 \\
            0.603492 & 0.487215 & -0.631205\\
          \end{array}
        \right)
\end{equation}

Table 2 also exhibits  intervals of the admissible K meson mass
corresponding to these solutions. Its changes under variations of
$\Delta{a}$ are in all domains relatively small and the ranges of
admissible values in different domains are strictly separated.

\begin{table}
 \centering
 \caption{The range of changes of the glueball contents under
 variation of $a$
 within the domains A, B, C and the octet content
 within the domain D. The intervals of the K-meson mass allowed over
 these domains are also shown. All masses are in GeV.}\label{2}
  \smallskip
\begin{tabular}{|c|c|c|c|}
               \hline
               A & $a\in((1.294)^2, (1.318)^2)$ & $1\geq(V_{23})^2\geq(0.764)^2$ & $1.388\leq m_K\leq 1.390$ \\
               B & $a\in((1.294)^2, (1.318)^2)$ & $1\geq(V_{33})^2\geq(0.842)^2$ & $1.353\leq m_K\leq 1.374$ \\
               C & $a\in((1.377)^2, (1.410)^2)$ & $1\geq(V_{13})^2\geq(0.810)^2$ & $1.420\leq m_K\leq 1.443$ \\
               D & $a\in((1.377)^2, (1.410)^2)$ & $1\geq   l_2^2  \geq(0.909)^2$ & $1.403\leq m_K\leq 1.410$ \\
               \hline
             \end{tabular}
\end{table}
It follows that the states of the pseudoscalar mesons $\pi(1300)$,
$K(1460)$, $\eta_1$, $\eta_2$, $\eta_3$ may constitute solution of
the ME. The price to pay for ignorance concerning both $\pi(1300)$
and $K(1460)$ meson masses is the ambiguity of the solution: instead
of unique solution we have four qualitatively different
one-parameter families of solutions. These families are defined
within four separated domains A, B, C, D of the $(a,b)$ plane and
can be distinguished due to the fact that the isoscalar physical
states are dominated by one of the $N$, $S$, $G$ or $\eta_8$
component. Hence the domination pattern enables us to distinguish
between the families of the solutions A, B, C, D of the ME. With the
present data on masses of the $\pi(1300)$ and $K(1460)$ mesons we
can only distinguish between the families. If one of these masses
was known then there would be only two solutions (not two families
of solutions !) as can be seen from the Fig. 1.

The data on flavor properties of the isoscalar mesons indicate the
family A as the one which points out the meson $\eta_2$ as a
particle dominated by the glueball state.\\

\section{Comments on solutions of the ME}
$\bullet\bullet$ A decuplet of mesons is a multiplet such that the
octet isoscalar state $\eta_8$ contributes to three isoscalar
physical states $\eta_i$. The contributions of the isoscalar octet
state to the physical states $\eta_i$, given by $l_i^2$s
(\ref{2.5}), constitute a solution of the ME. The coefficients $l_i$
are real numbers, therefore, the following conditions should be
satisfied
\begin{equation}\label{2.26a}
l_i^2>0 \quad /i=1,2,3/.
\end{equation}
This property is not guaranteed by solution (\ref{2.5}). Requiring
it, we put constraints on the masses of the decuplet.

The knowledge of the octet contents $l_i^2$s provides very
convenient way for constructing the mixing matrix.

This 3 x 3 orthogonal matrix can be parametrized by Euler angles.
The absolute values of trigonometric functions of two of the angles
can be expressed explicitly by $l_i^2$s, i.e. by the masses. The
requirement of the glueball flavor independence
\begin{equation}\label{2.26b}
    <G|m^2|u\bar u>=<G|m^2|d\bar d>=<G|m^2|s\bar s>,\\
\end{equation}
relates them to the trigonometric functions of the third angle.
However, some signs of the trigonometric functions cannot be
determined if only masses are known. To find them we use available
information on domination of the $\eta_i$ states by one of the $N$,
$S$, $G$, $\eta_8$ states. As a result, all the Euler angles, and
consequently, all elements of the mixing matrix are uniquely
determined \cite{enc}.\\

$\bullet\bullet$ The VEC description of the decuplet depends on the
number of ME (\ref{2.1}) which are taken into consideration.
There are two kinds of decuplets \cite{enc}.\\

i) A decuplet which is based on the assumption that three exotic
commutators vanish. Then four ME arise. If they were applied to the
nonet, they would define the ideal (I) one. We may imagine the
isoscalar states as superpositions of a glueball and the I nonet
states. We say that this decuplet is of the kind I. It complies with
one mass formula (\ref{2.6}) which, together with conditions
(\ref{2.26a}), defines explicit MOR restrictions for the masses
(\ref{2.9}). To also admit the degenerate solutions of the ME one
must allow $l_i^2\geq0$ for some $"i"$ and in the MOR $"\leq"$
instead of the $"<"$. The solutions of the ME do not include free
parameters -- all predicted quantities and relations are expressed
by physical masses.\\

ii) A decuplet arising under assumption that two exotic commutators
vanish. Then there are three ME. If applied to the nonet, they give
the S one. The decuplet is of the kind S if it is formed as
superposition of the glueball and the S nonet. In this case the mass
formula and ordering rule do not arise. The restrictions on the
masses are not so strict and follow from the conditions
(\ref{2.26a}).

If the masses of ten mesons with proper quantum numbers are known
and satisfy (\ref{2.26a}), but do not satisfy the MF, the decuplet
is of the kind S. If one of the masses is unknown then we can
determine it from the MF and the decuplet becomes of the kind I,
provided the masses satisfy (\ref{2.26a}). However, the states
constituting these two distinct decuplets may be not very different.

$\bullet\bullet$ The solutions A, B, C, D described above concern
the decuplet I. We now summarize the main features of the solution.

The analysis of the pseudoscalar decuplet presented here does not
give the unique result due to the fact that for determining two
unknown masses we have only one MF. Moreover, the MF is represented
by polynomial of the third degree with respect to each of these
variables.

The masses of $\pi(1300)$ and $K(1460)$ mesons are unknown. So the
values $a$ and $b$ which are natural variables in the VEC model are
also unknown, but they are bounded by MOR from the below and above.
These restrictions are helpful in choosing the proper solution of
the nonlinear equation (\ref{2.6}). Further restriction is provided
by the mass formula which cuts out the central part of the
MOR-allowed interval of $a$ and thus reduce it to two narrow
disconnected subintervals (see figure 1). To each $a$ belonging to
them there correspond two values of $b$. We divide the interval
$(x_2, x_3)$ of the values $b$ to two parts. As a result, the whole
domain of the values of the $a$ and $b$ is reduced to four small
domains A, B, C, D which are shown in figure 2. In these domains the
solution is unique if one of the variables, $a$ or $b$ is known.

Hence, due to the restrictions of the ME on $a$ and $b$, the
solution is split into four one-parameter, qualitatively different
families. The partition allows us to look for a solution in each
domain separately. Still we have two unknown masses and only one MF
equation relating them, but the domains are small and the solutions
are only slightly changing across them. The ranges of the masses of
the $a$ and $K$ mesons over the domains can be found out from the
table 2.

The partition of the whole domain of variables $(a, b)$ is
especially helpful for describing properties of mixing matrix. To
each of the domains A, B, C, D there is attributed a separate
one-parameter family of solutions of the MF determining the decuplet
-- among them the degenerate one. A family of the MF solutions, in
turn, induce one-parameter family of mixing matrices. All the
matrices of the family preserve a common dominance pattern. This
pattern is determined by the dominance of one of the N, S, G,
$\eta_8$ amplitudes in the  $\eta_1$, $\eta_2$, $\eta_3$ states and
can be read out from the degenerate solution. In each domain the
degenerate decuplet corresponds to a point at the outer corner of
the domain (see figure 2). Across the domain the \textit{pure} state
of the degenerate decuplet is transformed into \textit{dominating}
one and all isoscalar states become mixed. Within the domains A, B,
C the glueball dominates $\eta_2$, $\eta_3$, $\eta_1$ states,
respectively.

In the domain D the dominance pattern is different. The degenerate
decuplet consists of the octet of the exact symmetry and two
separate singlets being mixed states of the $(q\bar{q})_{singlet}$
and $G$. $\eta_2$ is the pure octet $\eta_8$, while $\eta_1$ and
$\eta_3$ are pure singlets. The rates of the $(q\bar{q})_{singlet}$
and $G$ states in the structures of the $\eta_1$ and $\eta_3$ mesons
are comparable, slowly changing functions of the parameter
$\Delta{a}$ within the domain. In spite of the identical flavor
properties of the constituents, the properties of the $\eta_1$ and
$\eta_3$ mesons should be different and the difference is changing
across the domain. This is mainly due to the fact that they are
opposite superpositions of the $G$ and $(q\bar{q})_{singlet}$
amplitudes. In this family of solutions the glueball state is not
apparent.

On account of so different properties of the families A, B, C, D the
qualitative information on the isoscalar mesons is sufficient to
make the choice. The proper family can be chosen on the basis of the
flavor properties of the isoscalar mesons $\eta_1$, $\eta_2$,
$\eta_3$. As it has been pointed out, even if all masses are known
and satisfy MF and MOR, such an extra information is necessary for
constructing the mixing matrix. An exact solution of the ME
corresponding to definite values of the $\pi(1300)$ and K(1460)
masses would be determined by suitable value of $\Delta{a}$.

$\bullet\bullet$ The restrictions following from figure 1 do not
hold for the decuplet of the type S. There is no mass formula in
this case; therefore, there is no connection between the variables
$a$ and $b$ and there is no mass gap (${x_P, x_Q}$) of the $a$
meson. Also the MOR (\ref{2.9}) does not exist. Such a situation can
arise for the decuplet we discuss if the measured masses of the
$\pi(1300)$ and K(1460) mesons will not satisfy MF. However, to
define a decuplet of any type we always need mesons having such
masses that conditions (\ref{2.26a}) are satisfied. These conditions
give weaker constraints on $a$ and $b$. We find
\begin{equation}\label{2.26c}
    x_1<x_8<x_3,
\end{equation}
where $x_8$ is given by the GMO mass formula,
\begin{equation}\label{2.26d}
    x_8=\frac{1}{3}a+\frac{2}{3}b.
\end{equation}

The glueball contents of the isoscalar mesons $\eta_1$, $\eta_2$,
$\eta_3$ can be always calculated from (\ref{2.5}) if the masses
$m_\pi$ and $m_K$ satisfy $l_i^2>0$ /i=1,2,3/.

Having known the $l_i^2$s we can construct the mixing matrix. If the
state of $\eta_2$ is predicted to be dominated by $G$ the solution
of the ME should be similar to the one describing the states of the
family A.

\section{Pseudoscalar versus scalar meson multiplets}
\subsection{Parity related spectra of the spin 0 mesons}
Having described the multiplets of pseudoscalar mesons we get the
opportunity to confront its properties with the properties of the
corresponding multiplets of scalar mesons \cite{enc}. The comparison
may reveal some new features of the meson spectroscopy.

Let us compare the $0^{-+}$ and $0^{++}$ multiplets.\\
The ground states form the nonets:
\begin{equation}\label{2.28}
    \pi, \quad K, \quad\eta,\quad  \eta',
\end{equation}
\begin{equation}\label{2.29}
a_0(980),\quad K_0(1430),\quad f_0(980),\quad f_0(1710).
\end{equation}
which are followed by the decuplets:
\begin{equation}\label{2.30}
    \pi(1300),\quad K(1460),\quad \eta(1295),\quad \eta(1405),\quad
\eta(1475),
\end{equation}
\begin{equation}\label{2.31}
a_0(1450),\quad K_0(1950),\quad f_0(1370),\quad f_0(1500),\quad
f_0(2200)/f_0(2330).
\end{equation}
In both cases we have the same sequence of the multiplets.  Some of
the masses are not exactly known, but this does not spoil the
general picture.

Let us observe that not only the sequences of the multiplets are
similar, but also the inner structures of the decuplets are; namely: \\
- the physical mesons $f_0(1500)$ and $\eta(1405)$ which are
dominated by glueball states are settled just between the remaining
isoscalars which are expected to be mostly the $N$ and $S$ quark states,\\
- both decuplets involve excited $q\bar{q}$ states; hence both
glueballs mix with the excited ${(q\bar{q})_{isoscalar}}$ states.

The latter property suggests affinity of the glueball with the
excited states. This is especially prompted by the mixing of the
$0^{++}$ glueball. Its mass belongs to the region where the nonet
ground states and the decuplet excited states are overlapping, but
the glueball prefers mixing just with the excited $q\bar{q}$ states
-- there is no trace of mixing with the ground $q\bar{q}$ states
\cite{enc}.

The $0^{-+}$ and $0^{++}$ mesons form the parity related spectra of
multiplets (nonets and decuplets). The sequences of these multiplets
differ only due to existence of the scalar meson $\sigma(600)$ which
has no adequate pseudoscalar partner. But the nature of this meson
is still a matter of discussion. Several authors suggest that its
nature is different from the nature of other mesons \cite{Pen,
WOchs, Anisov}. By ignoring the $\sigma(600)$ we find that $0^{-+}$
and $0^{++}$ mesons form parity related {\it spectra of multiplets}.

The transparency of this picture confirms not only the opinion about
the distinct nature of the $\sigma(600)$, but also supports
correctness of sorting out the scalar mesons between the overlapping
multiplets \cite{enc}.

However, there is also a difference between these spectra. The mass
spread of the $0^{-+}$ multiplets is shrinking for consecutive
multiplets (nonet $137\div958\ \mathrm{MeV}$, decuplet
$1295\div1475\, \mathrm{MeV}$, perhaps degenerate octet at 1800
MeV). The tendency of shrinking the mass spread of the higher lying
multiplets is even more clearly expressed in the spectrum of
$1^{--}$ mesons where all known multiplets above 1400 MeV (at 1400
MeV and 1800 MeV) are degenerate octets \cite{Sum}. But this
tendency is not seen in the spectrum of the $0^{++}$ multiplets - at
least below 2300 MeV.

\subsection{The masses of the spin 0 glueballs}
The LQCD calculations predict for the lower bound of the lightest
$0^{++}$ glueball the mass \cite{Bali, Mor, Che}
\begin{equation}\label{2.311}
    m_{G^{++}}\approx1.500 GeV.
\end{equation}
Such mass allows to attribute the glueball nature to several mesons
-- among them to favored $f_0(1500)$.

For the lightest $0^{-+}$ glueball these calculations predict the
lower bound at
\begin{equation}\label{2.312}
    m_{G^{-+}}\approx2.300 GeV.
\end{equation}
With this value no isoscalar meson discussed here can be assigned to
be the glueball. Attempts to diminish this bound were unsuccessful.

The LQCD calculations also predict the lower bounds for the masses
of many other glueballs with different $J^{PC}$. The result is that
the mass (\ref{2.311}) marks the minimum of these lower bounds.
However, this result can be obtained also in other approaches (see,
e.g. West's theorem \cite{West}). Therefore, it is considered more
general and independent of particular approach.

The LQCD predicts masses of pure glueball states. Also the flux tube
(FT) \cite{Fad} concerns such states.

The FT approach predicts, however, that masses of the $0^{-+}$ and
$0^{++}$ glueballs should be equal. Since $f_0(1500)$ is the favored
$0^{++}$ glueball candidate, we should expect the $0^{-+}$ glueball
mass at about $1.500 GeV$. Hence, the LQCD and FT predictions on
$0^{-+}$ glueball mass are contradictory.

The VEC prediction of the glueball mass has different source. It
refers to broken unitary symmetry which collects the mesons in
multiplets -- the octets and the nonets. Several of them are well
established in the low mass region at various $J^{PC}$. We assume
that at higher masses the mesons are collected in multiplets as
well. In the case where the glueball appears we expect the decuplet.
The three isoscalar components of the decuplet are superpositions of
the $q\bar{q}$ and $G$ states. There is no pure glueball state but
such state may dominate one of the isoscalars.

The mass formula for the decuplet relates physical masses. Also the
mixing matrix is explicitly determined by physical masses. There is
no ambiguity and only physical masses enter. Therefore, predictions
are definite and, in favorable case, may help to perceive something
new.

Using the mixing matrix we can calculate the mass of the pure
glueball state. We can do this for the decuplets $0^{-+}$ and
$0^{++}$ separately (not assuming any relation between them).

From the decuplet $0^{-+}$, assuming solution A and the mass input
appropriate to the mixing matrix (\ref{2.23}) we find
\begin{equation}\label{2.313}
m_{G^{-+}}=1.369 GeV.
\end{equation}

From the decuplet $0^{++}$ for solution 1 \cite{enc} we get
\begin{equation}\label{2.314}
    m_{G^{++}}=1.497 GeV.
\end{equation}

The difference between these predictions is approximately equal to
the $\pi$ meson mass.
\begin{equation}\label{2.315}
    m_{G^{++}}-m_{G^{-+}}=m_\pi.
\end{equation}
Also observe that the inequality
\begin{equation}\label{2.316}
    m_{G^{-+}}<m_{G^{++}}
\end{equation}
holds for all families A, B, C, D despite of LQCD calculations and
West's Theorem predictions.

Let us comment.

The VEC search of the glueball is carried on within the isoscalar
sector of the decuplet. The prediction of the glueball mass consists
in setting all masses of the decuplet and fitting the mixings of the
isoscalar components to their flavor properties. Hence, the mixings
play an important role in glueball determination.

The FT prediction of the equality of the $0^{-+}$ and $0^{++}$
glueball masses is approximately obeyed. This may follow from the
fact that
$G$ contributions to their structure are high and almost equal:\\
for $0^{-+}$ decuplet (solution A (\ref{2.23})),
\begin{equation}\label{2.317}
    V_A=
          \begin{bmatrix}
            xxxxxx & xxxxxx & xxxxxx \\
            xxxxxx & xxxxxx & -0.94154 \\
            xxxxxx & xxxxxx & xxxxxx \\
          \end{bmatrix}
        .
 \end{equation}
for the $0^{++}$ decuplet \cite{enc},
\begin{equation}\label{2.318}
    V_1=
    \begin{bmatrix}
      xxxxxx & xxxxxx & xxxxxx \\
      xxxxxx & xxxxxx & -0.88466 \\
      xxxxxx & xxxxxx & xxxxxx \\
    \end{bmatrix}.
\end{equation}

In fact, the masses of the $0^{-+}$ and $0^{++}$ glueballs are not
identical. Difference between them, although small in the scale of
the mass of these glueballs, is not negligible. This difference is a
result of the parallel but independent examination of the relations
between physical masses of the decuplet particles. Perhaps,
different masses of these glueballs suggest that different decuplets
affect the glueball component nonidentically. That may concern not
only the cases of different $J^{PC}$ but also decuplets of the same
$J^{PC}$ in different mass regions (if such ones exist).

The West's theorem is not fulfilled: the observed mass difference
(left part of the (\ref{2.315})) has opposite sign. The absolute
value of the difference is not predicted. Also the difference
between the masses predicted by LQCD has wrong sign. Beside, the
value of this difference is probably too large to be explained as
the mixing effect.

\section{Conclusions}
1. Owing to the unknown masses of the $\pi(1300)$ and $K(1460)$
mesons the solution of the ME for decuplet $\pi(1300)$, $K(1460)$,
$\eta(1295)$, $\eta(1405)$, $\eta(1475)$ is not unique. In spite of
that due to the restrictions of the VEC model all solutions can be
classified into four separate families; one of the families points
out $\eta(1405)$ as the particle dominated by the glueball state.\\
2. The scalar and pseudoscalar glueballs belong to the decuplets
formed by mixing $G$ with excited $q\bar q$ isoscalar states.\\
3. The spectra of the known multiplets of the $0^{-+}$ and $0^{++}$
mesons are parity related provided the $\sigma(600)$ is ignored.\\
4. The mass of the pure pseudoscalar glueball state $m_{G^{-+}}$ is
smaller than the mass $m_{G^{++}}$ of the scalar glueball one.\\

\section{Acknowledgments}
The authors thank Professors P. Maslanka, J. Rembielinski, W. Tybor
and management of BLTF JINR for promoting our cooperation as well as
Dr K. Smolinski for help in computer operations. Especially we thank
Professor S. B. Gerasimov for many valuable discussions in early
stage of this work and Professor P. Kosinski for many interesting
comments. This work was financially supported by JINR B-I Fond and
by grants nos690 and 795 of University of Lodz.

\newpage

\section{Appendix. Solving the MF}

Introduce in each of the domains A, B, C, D shown in figure 2 the
small variables $\Delta{a}$ and $\Delta{b}$. They can be chosen to
be nonnegative $(\Delta{a}, \Delta{b}\geq{0})$. Therefore, in the
individual domains A, B, C, D, we have:

\begin{figure}
\centering
\includegraphics[angle=-90,width=\textwidth]{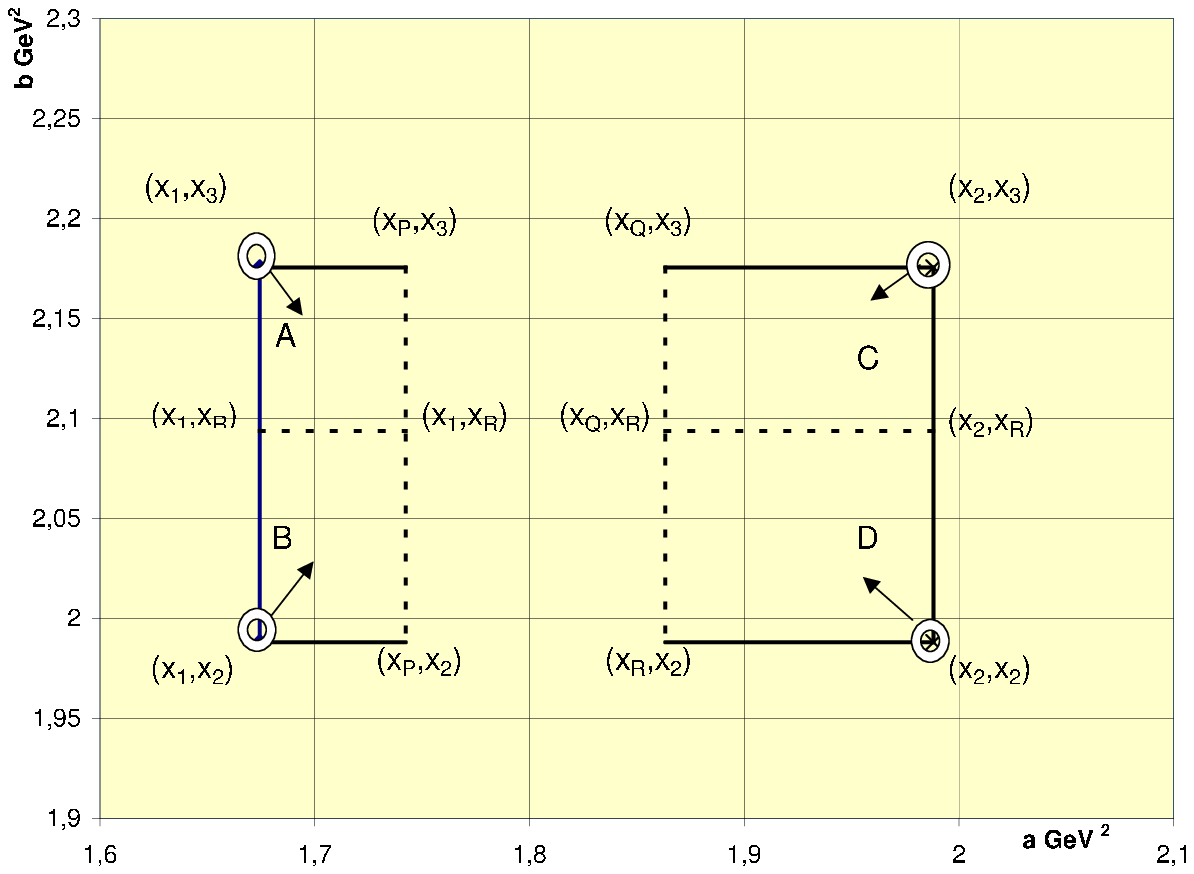}
\caption{The domains A, B, C, D of the solutions of MF. The double
circles at the outer corners of all domains correspond to degenerate
decuplets. In each domain the pairs of values $(a,b)$, being the
solutions of the MF, lie on a curve connecting this corner with the
opposite one. The curves are close to diagonals of the domains. The
arrows beginning at the points of degeneracy indicate directions
towards the growing mixing of the states.} \label{Figure1A}
\end{figure}

\begin{subequations}\label{2.15}
\begin{align}
    \label{2.15a}
    A:&&   a&=x_1+\Delta{a},&    b&=x_3-\Delta{b},\\
    \label{2.15b}
    B:&&   a&=x_1+\Delta{a},&    b&=x_2+\Delta{b},\\
    \label{2.15c}
    C:&&   a&=x_2-\Delta{a},&    b&=x_3-\Delta{b},\\
    \label{2.15d}
    D:&&   a&=x_2-\Delta{a},&    b&=x_2+\Delta{b}.
\end{align}
\end{subequations}

Further procedure is the following. Substituting $a$ and $b$ into
(\ref{2.5}), we express the coefficients $l_i^2$ as functions of the
$\Delta{a}$ and $\Delta{b}$. Putting $a$ and $b$ into MF
(\ref{2.15}) we get the relation between the  $\Delta{b}$ and
$\Delta{a}$. This relation is a cubic equation with respect to any
of these variables. For our purposes it is sufficient to find the
approximate solution. We have
\begin{equation}\label{2.16}
    \Delta{b}=\Delta{b}(\Delta{a},x_1, x_2, x_3).
\end{equation}
Substituting this function into (\ref{2.5}) we obtain the functions
$l_i^2(\Delta{a},x_1, x_2, x_3)$. If all these functions are
positive, we may consider the corresponding values of $a$ and $b$,
together with the functions $l_i^2(\Delta{a},x_1, x_2, x_3)$, as an
approximate solution of the ME (\ref{2.1}) in the appropriate
domain. This procedure is to be performed for all the domains A, B,
C, D.

In the domains A and B we may neglect all terms of (\ref{2.6})
containing higher degrees of $\Delta{a}$ or $\Delta{b}$ and restrict
ourselves to the linear dependence between them. The approximation
is plausible for $\Delta{a}$ covering all the interval $(x_1,x_P)$.
We obtain
\begin{subequations}\label{2.17}
\begin{align}
    \label{2.17a}
    A:&&   \Delta{b}&=\frac{\Delta{a}}{2}\frac{x_2-x_1}{x_3-x_1},\\
    \label{2.17b}
    B:&&   \Delta{b}&=\frac{\Delta{a}}{2}\frac{x_3-x_1}{x_3-x_2}.
\end{align}
\end{subequations}
In the domains C and D we also take into account the term quadratic
in $\Delta{b}$ and all powers of $\Delta{a}$. This is to avoid
$l_1^2<0$ in the domain C and to extend applicability of this
approximation towards the largest values of $\Delta{a}$ in the
domain D. In these cases the expressions for $\Delta{b}$ are the
solutions of the quadratic equation, so they are simple but long and
we do not write them out. Two solutions of the quadratic equation
for $\Delta{b}$ do not cause confusion, as only one of them complies
with the condition $l_i^2>0$ for all i=1,2,3. In both regions the
approximation is plausible everywhere, except the small surroundings
of the point $b=x_R$.

In all the solutions A, B, C, D the value $\Delta{a}=0$ implies
$\Delta{b}=0$. The degeneracy of the decuplet is destroyed if
$\Delta{a}>0$. An isoscalar state $\eta_i$ having pure $G$ or pure
$\eta_8$ structure becomes mixed. However, it is still dominated by
the same state provided $\Delta{a}$ is sufficiently small. The
mixing is intensified and the dominance is getting weaker as
$\Delta{a}$ is increasing. By examining the mixing matrix we can
check whether the dominance is kept inside all domains.

Table 2 shows the range of change of the squared matrix elements
$V_{23}$, $V_{33}$, $V_{13}$ expressing contribution of the glueball
to the $\eta_2$, $\eta_3$, $\eta_1$ respectively and the octet
content $l_2^2$ under change of $\Delta{a}$ in the solutions A, B,
C, D. It can be seen that dominance of the $G$ and $\eta_8$ states
is kept over the whole domain of these solutions. The $N$ and $S$
dominance of the other $\eta_i$ states belonging to the same
solution (not shown in the table) is preserved across the domains A,
B, C as well; however, there is no dominance of $\eta_1, \eta_2,
\eta_3$ by $N, S, G$ in the case of the solution D. We thus find
that within each domain the solution has specific \textit{dominance
pattern} which does not change under variation of $\Delta{a}$.
(Obviously, the degree of the dominance does depend on the
$\Delta{a}$). The dominance patterns of the solutions A, B, etc.,
are identical with the patterns of ideal structures of the
degenerate decuplet (\ref{2.19}). These structures correspond to the
points at the outer corners of the appropriate domain.

To conclude, all solutions of the ME are split into four separate
one-parameter families. The solutions belonging to the same family
are slightly different. The solutions belonging to different
families have different dominance patterns.

\end{document}